\documentclass[letter]{aa}  
\usepackage{graphicx}
\usepackage{txfonts}

\usepackage{inputenc}

\begin{document}

   \title{Formation of compact galaxies %at $z \sim 2$
   in the Extreme-Horizon simulation}% (2)The Extreme-Horizon simulation: compact galaxies from gas accretion and repeated major mergers of low-mass galaxies, (1) important role of repeated tiny mergers in the compaction of high-redshift galaxies in the Extreme-Horizon simulation}
\titlerunning{Compact galaxies in the Extreme-Horizon simulation}

   \author{S. Chabanier          \inst{1,2}
          \and
          F. Bournaud\inst{2,1}
          \and
          Y. Dubois\inst{3}
          \and
          S. Codis\inst{3,4}
          \and
          D. Chapon\inst{1}
          \and
          D. Elbaz \inst{2}
          \and
          C. Pichon \inst{3,4,5}
          \and          
          O. Bressand \inst{6,7}
          \and
          J.~Devriendt \inst{8}
          \and
          R. Gavazzi \inst{3}
          \and
          K. Kraljic \inst{9}
          \and
          T. Kimm \inst{10}
          \and
          C. Laigle \inst{3}
          \and
          J.-B. Lekien \inst{6,7}
          \and
          G. W. Martin \inst{11}
          \and
          N.~Palanque-Delabrouille\inst{1}
          \and 
          S. Peirani \inst{12}
          \and 
          P.-F. Piserchia \inst{6,7}
          \and
          A. Slyz\inst{13}
          \and
          M. Trebitsch\inst{14,15}
          \and
          C. Y\`eche\inst{1}
          }

\authorrunning{S. Chabanier et al.}

   \institute{IRFU, CEA, Universit\'e Paris-Saclay, 91191 Gif-sur-Yvette, France
         \and
         AIM, CEA, CNRS, Universit\'e Paris-Saclay, Universit\'e Paris Diderot, Sorbonne Paris Cit\'e, 91191 Gif-sur-Yvette, France
              \and
         CNRS and Sorbonne Universit\'e, UMR 7095, Institut d'Astrophysique de Paris, 98 bis Boulevard Arago, 75014 Paris, France
         \and
         Universit\'e Paris-Saclay, CNRS, CEA, Institut de Physique Th\'eorique, 91191 Gif-sur-Yvette, France
         \and
         Korea Institute of Advanced Studies (KIAS) 85 Hoegiro, Dongdaemun-gu, Seoul, 02455, Republic of Korea
         \and
         CEA, DAM, DIF, F-91297 Arpajon, France
         \and
         Université Paris-Saclay, CEA, Laboratoire en Informatique Haute Performance pour le Calcul et la simulation, 91680 Bruyères-le-Châtel, France
         \and
         Astrophysics, University of Oxford, Oxford OX1 3RH, UK
         \and
         Institute for Astronomy, Royal Observatory, Edinburgh EH9 3HJ, United Kingdom
         \and
         Department of Astronomy, Yonsei University, 50 Yonsei-ro, Seodaemun-gu, Seoul 03722, Republic of Korea
         \and
         Steward Observatory, University of Arizona, 933 N. Cherry Ave, Tucson, AZ 85719, USA
         \and
         Universit\'e Côte d’Azur, Observatoire de la Côte d’Azur, CNRS, Laboratoire Lagrange,CS 34229, 06304 Nice Cedex 4, France
         \and
         Astrophysics, University of Oxford, Denys Wilkinson Building, Keble Road, Oxford OX1 3RH, UK
         \and
         Max-Planck-Institut f{\"u}r Astronomie, K{\"o}nigstuhl 17, 69117 Heidelberg, Germany
         \and
         Zentrum f{\"u}r Astronomie der Universit{\"a}t Heidelberg, Institut f{\"u}r Theoretische Astrophysik, Albert-Ueberle-Str. 2, 69120 Heidelberg, Germany
             }

   \date{Received September 15, 20xx; accepted March 16, 20xx}

 \abstract{We present the Extreme-Horizon (EH) cosmological simulation: EH models galaxy formation with stellar and AGN feedback and uses a very high resolution in the intergalactic and circumgalactic medium.
The high resolution in low-density regions results in smaller-size massive galaxies at redshift $z=2$, in better agreement with observations compared to other simulations. This results from the improved modeling of cold gas flows accreting onto galaxies. Besides, the EH simulation forms a population of particularly compact galaxies with stellar masses of $10^{10-11}$\,M$_\sun$ that are reminiscent of observed ultracompact galaxies at $z\simeq2$. These objects form mainly through repeated major mergers of low-mass progenitors, independently of baryonic feedback mechanisms. This formation process can be missed in simulations using a too low resolution in low-density intergalactic regions.
}%{}{}{}{} 

   \maketitle
%
%-------------------------------------------------------------------
\section{Introduction}

Early-type galaxies (ETGs) at redshift $z>1.5$ are much more compact than nearby ones \citep{daddi05}. At stellar masses about $10^{11}$\,M$_\sun$, they typically have half-mass radii of 0.7--3 kpc, about three times smaller than nearby ellipticals with similar masses \citep{vanderwel}. Compact radii come along with steep luminosity profiles and high Sersic indices \citep[][]{vdk-brammer2010,carollo13}. Star-forming galaxies (SFGs) also decrease in size with increasing redshift \citep[e.g.,][]{kriek09, dutton-vdb11}. Besides, the CANDELS survey has discovered a population of very compact SFGs at z$\sim$2: the so-called ``blue nuggets'' \citep{barro13,williams-giavalisco14} have stellar masses of $10^{10-11}$\,M$_\sun$ with unusually small effective radii around 2\,kpc and sometimes even below 1\,kpc. Compact SFGs have high comoving densities, about $10^{-4}$\,Mpc$^{-3}$ for stellar masses above $10^{10}$\, M$_\sun$, and $10^{-5}$\,Mpc$^{-3}$ above $10^{11}$\,M$_\sun$ \citep{wang19}. In addition, SFGs at $z\simeq 2$ often have very compact gas and star formation distributions \citep{elbaz18}.

Many  processes have been proposed to explain the formation of compact galaxies, ranging from early formation in a compact Universe \citep{lilly16} to the compaction of initially-extended galaxies \citep{zolotov15} through processes that may include galaxy mergers, disk instabilities \citep{BEE07,dekel-burkert14}, triaxial haloes \citep{tomasseti16}, accretion of counter-rotating gas \citep{danovich} or gas return from a low-angular momentum fountain \citep{elm14}.

The Extreme-Horizon (EH) cosmological simulation, presented in Sect.~2, models galaxy-formation processes with the same approach as Horizon-AGN \citep[HAGN,][hereafter D14]{Dubois2014} and a substantially increased resolution in the intergalactic and circumgalactic medium (IGM and CGM). The properties of massive galaxies in EH and the origin of their compactness are studied in Sect.~3 and~4.

%-------------------------------------------------------------------
\section{The Extreme-Horizon simulation}

  \begin{table*}
  \begin{tabular}{c|cccccccc}
   \hline \hline
  grid resolution [$\rm kpc.h^{-1}$] & 100 & 50 & 25 & 12.5 & 6.25 & 3.12 & 1.56 & 0.78 \\
  \hline
   $\rho_{\rm DM, thresh}/\rho_{\rm DM, mean}$ (EH) & -- & min resolution & 1.3 & 10 & 82 & 655 & 26,340 & 210,725\\

    $\rho_{\rm DM, thresh}/\rho_{\rm DM, mean}$ (SH)   & min resolution & 6.4 & 51.2 &410 & 3,277 & 26,214 & 210,725 & 1,685,800 \\
    \hline
    volume fraction (EH) & -- & 45\% & 43\% & 10\% & 1\% & 0.04\% & -- & -- \\
     volume fraction (SH) & 80\% & 17\% & 2\% & 0.17 \% & 0.013\% & $5\times10^{-4}$\% & -- & -- \\
     volume fraction (HAGN) & 77\% & 19\% & 2\% & 0.2 \% & 0.01\% & $6\times10^{-4}$\% & -- & -- \\
     \hline
 \end{tabular}
 \label{tab:reso}
 \caption{Resolution strategy for EH and SH: critical dark matter densities, in units of mean cosmological density, required to activate refinements are listed. The critical baryon densities vary in the same proportions for EH and SH. The volume fractions measured at each resolution level are given at $z$=2 and compared to HAGN.}
\end{table*} 

 The EH simulation is performed with the adaptive mesh refinement code RAMSES \citep{Teyssier2002} using the physical models from HAGN (D14). The spatial resolution in the CGM and IGM is largely increased compared to HAGN, while the resolution inside galaxies is identical, at the expense of a smaller box size of 50\,Mpc\,h$^{-1}$. The control simulation of the same box with a resolution similar to HAGN is called Standard-Horizon (SH). EH and SH share initial conditions realized with \textsf{mpgrafic} \citep{Prunet2008}. 
These use a $\rm \Lambda CDM$ cosmology with matter density $\mathrm{\Omega_m}$ = 0.272, dark energy density $\mathrm{\Omega_{\Lambda}}$ = 0.728, matter power spectrum amplitude $\sigma_8$ = 0.81, baryon density $\Omega_b$ = 0.0455, Hubble constant $\mathrm{H_0}$ = 70.4 km\,s$^{-1}$\,Mpc$^{-1}$, and scalar spectral index $\mathrm{n_s}$ = 0.967, based on the WMAP-7 cosmology \citep{Komatsu2011}. EH was performed on 25,000 cores of the AMD-Rome partition of the Joliot Curie supercomputer at TGCC
 and partly used the {\tt Hercule} parallel I/O library \citep{bressand, Strafella2020}.

\subsection{Resolution strategy}
\label{sec:reso}

SH uses a $512^3$ coarse grid, with a minimal resolution of 100\,kpc\,h$^{-1}$ as in HAGN. Cells are refined up to a resolution of $ \simeq$ 1\,kpc in a quasi-Lagrangian manner: any cell is refined if the dark matter (DM) mass and/or baryonic mass exceed eight times the initial DM mass or baryonic mass in coarse cells. This resolution strategy matches that of HAGN (Table~\ref{tab:reso}).

EH uses a $1024^3$ coarse grid and a more aggressive refinement strategy: the whole volume is resolved with a twice higher resolution and most of the mass is resolved with a four times higher resolution in 1-D, yielding an improvement of 8 to 64 for the 3-D resolution. This improvement continues until the highest resolution of $ \simeq$ 1\,kpc is reached: the critical densities to activate refinements are listed in Table~\ref{tab:reso}, which also indicates the volume fraction at various resolution levels. Such aggressive approach for grid refinement can better model the early collapse of structures \citep{Oshea2005}. Appendix~\ref{app:EH} illustrates the resolution achieved in representative regions of the CGM and IGM in EH and SH. The resolution in EH haloes is typically of 6 kpc while 25 kpc for SH. However, galaxies themselves are treated at the very same resolution in EH and SH: any gas denser than 0.1\,cm$^{-3}$ is resolved at the highest level in SH, as is also the case for 90\% of the stellar mass.

\subsection{Baryonic physics}
\label{sec:physics}

Like in HAGN (D14), reionization takes place after redshift 10 due to heating from a uniform UV background from \cite{Haardt1996}. H and He cooling are implemented as well as metal cooling following the \citet{Sutherland1993} model. 

Star formation occurs in cells with an hydrogen number density larger than $\mathrm{\rho_0 = 0.1 H/cm^{3}}$. The star formation rate density is $\dot { \rho } _ { * } = \epsilon _ { * } \rho / t _ { \mathrm { ff } }$ where $ t _ { \mathrm { ff } }$ is the local gas free-fall time and $\mathrm{ \epsilon _ { * } = 0.02}$ is the star formation efficiency \citep{Kennicutt1998}. Mass, energy and metals are released by stellar winds, type Ia and type II supernovae assuming a Salpeter Initial Mass Function. 

Black holes (BH) are represented by sink particles with an initial mass of $10^5$M$_\sun$. They accrete gas through an Eddington-limited Bondi-Hoyle-Lyttleton model. Boosted accretion episodes are included when the gas density overcomes a density threshold to mitigate resolution effects, the boosting being calibrated to produce realistic BH masses. The AGN feedback comes in two modes \citep{Dubois2012}: the quasar mode injects thermal energy and the radio mode injects mass, momentum and kinetic energy in the surrounding medium. We refer the reader to D14, the analysis of \cite{dubois16} and \cite{Dubois2012} for the detailed parameterization of these models.

%-------------------------------------------------------------------
\section{Galaxy compaction in EH}

\subsection{Galaxies in the EH simulation}
\label{sec:gal}

\begin{figure}[!h]
   \centering
   \includegraphics[height=2.9cm]{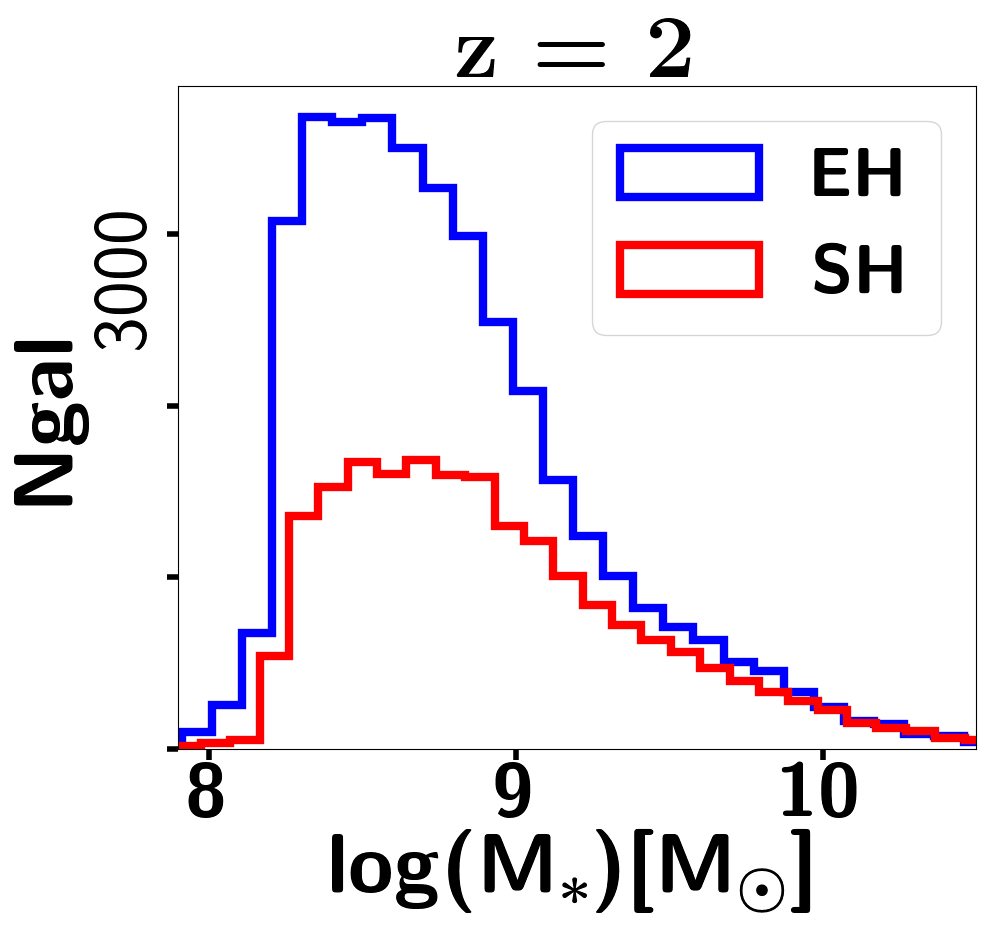}
   \includegraphics[height=2.9cm]{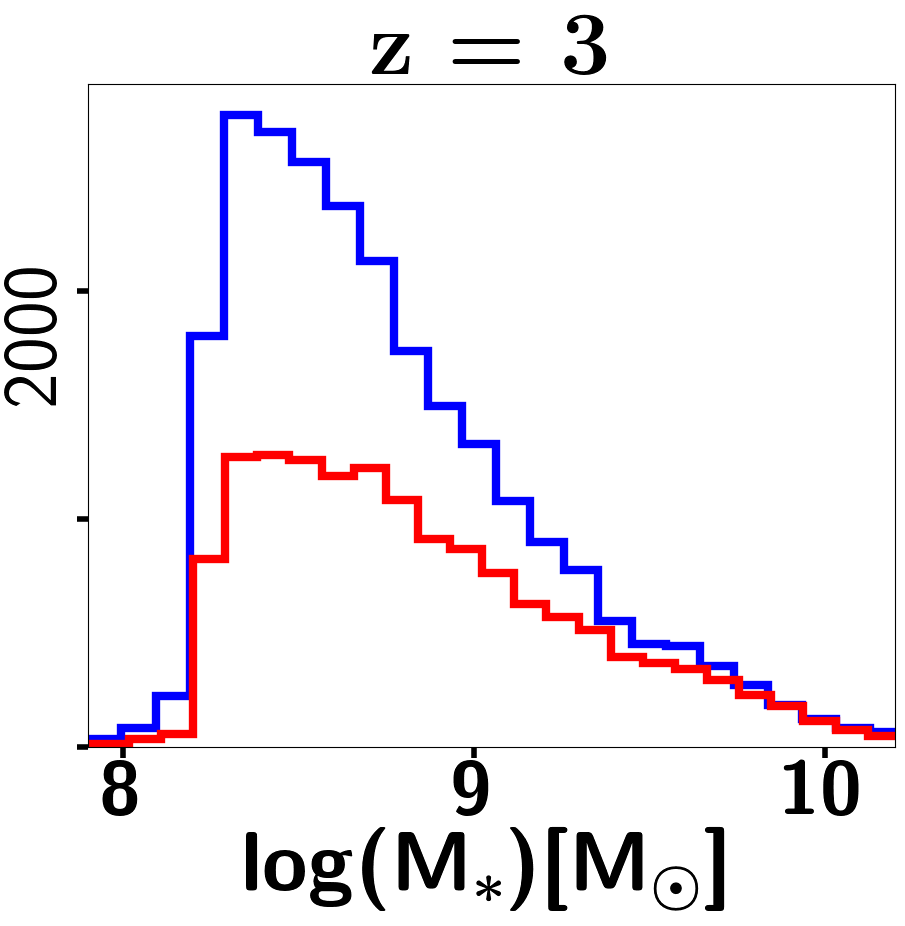}
   \includegraphics[height=2.9cm]{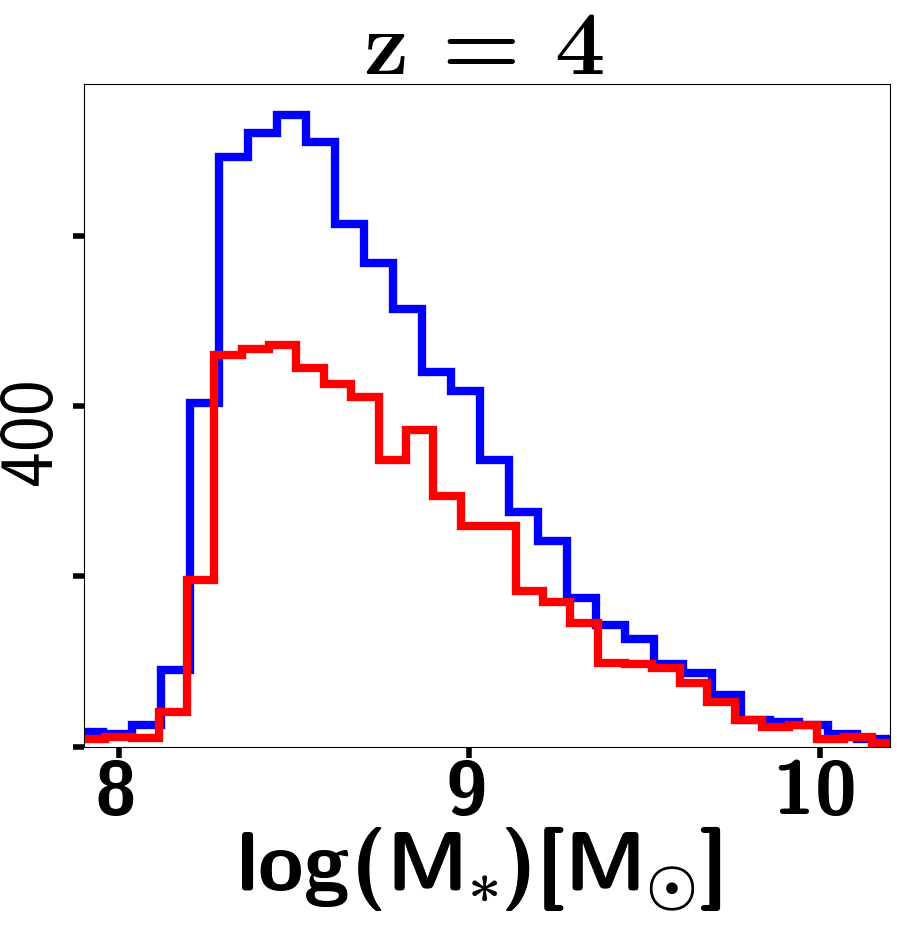}
      \caption{Number of galaxies per mass bin in EH and SH at $z$ = 2, 3 and 4.}
         \label{fig:Mgal}
   \end{figure}

We detect galaxies with more than 50 stellar particles (about $10^8$\,M$_\sun$) using AdaptaHOP \citep{aubert04}. 37,698 galaxies are detected in EH at $z \sim 2$ and 20,314 in SH, with stellar mass functions at various redshifts shown in Fig.~\ref{fig:Mgal}. While the mass functions above $10^{10}$\,M$_{\odot}$ are quite similar in both simulations, EH forms twice as many galaxies as SH with stellar $\mathrm{M_*} \leq 5 \times 10^9\,$M$_{\odot}$. We rule out any detection bias as stellar particles have similar masses in EH and SH, and attribute this difference to the increased resolution in low-density regions. Fitting the $z=2$ mass function with a power-law of the form $\Phi(M_*) \propto M_*^\beta$ in the $10^9\leq \log (M_*/M_\sun)\leq 10^{9.5}$ range yields $\beta = -0.68$ for EH and $-0.34$ for SH. Observations indicate a slope $-1.0\leq \beta\leq -0.5$ in this mass range \citep{santini2012,tomczak2014}, showing that low-mass galaxy formation is substantially under-resolved or delayed in SH.

We build samples of galaxies with $M_* \geq 5\times 10^{10}$\,M$_\sun$. On-going major mergers identified through the presence of a companion with more than 20\% of the stellar mass within 20\,kpc and/or a double nucleus, are rejected, yielding a sample of massive galaxies displayed in Appendix~\ref{app:maps} for each simulation.
We then study the mass distribution of the selected galaxies, taking into account non-sphericity. Stellar density maps are computed with a 500\,pc pixel size. Pixels below 50\,M$_\sun$\,pc$^{-2}$, typically corresponding to a surface brightness $\mu_i \geq 28\,\mathrm{mag\,arcsec}^{-2}$, are blanked out. Ellipse-fitting of iso-density contours is performed using the technique from \citet{krajnovic}.
Satellite galaxies are removed as follows: the circular region centered on the luminosity peak of the companion and extending up to the saddle of the luminosity profile between the main galaxy and the companion is ignored in the ellipse-fitting procedure, and replaced with the density profile modeled on other regions. Satellites with a mass below 5\% of the main galaxy are ignored to avoid removing sub-structures of the main galaxy.
Three perpendicular projections are analyzed for each galaxy, and the median results are kept for both the stellar mass $M_*$ and the half-mass radius $R_e$, the latter being defined as the semi-major axis of the isophote-fitting ellipse containing 50\% of the stellar mass. The removal of satellite galaxies and low-density outskirts yields final stellar masses slightly below the initial estimates, down to $M_*$\, $\simeq$\,$3-4$$\times$\,10$^{10}$\,M$_\sun$.
%FB

\begin{figure}[!h]
   \centering
   \includegraphics[width=\columnwidth]{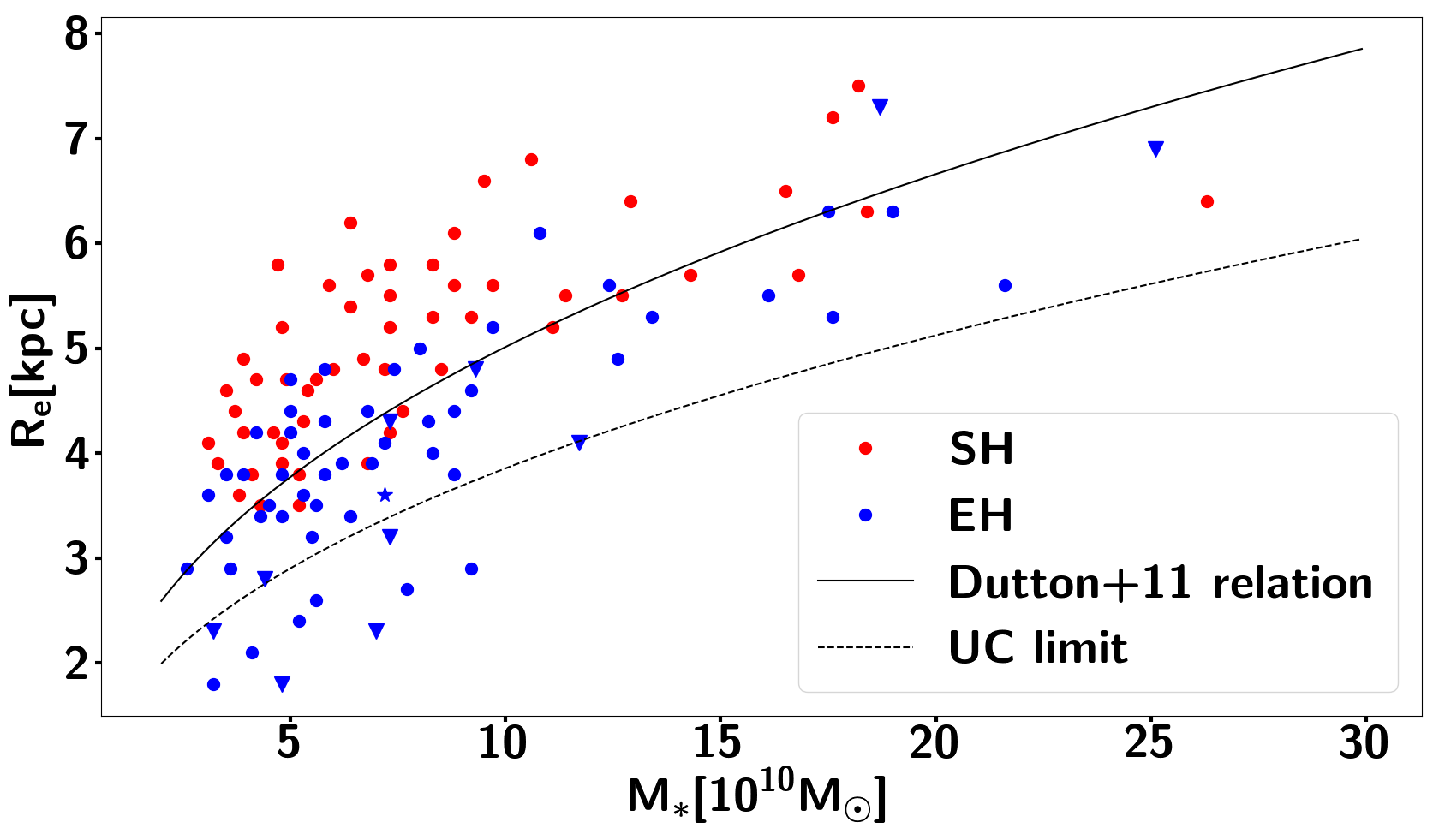}
      \caption{Stellar half-mass radius $R_e$ versus stellar mass $M_*$ for massive galaxies at $z=2$ in EH and SH. The displayed model from \citet{dutton-vdb11} provides a good fit to SFGs at $z$=2. UC galaxies lie below the black dashed line while NUC galaxies are above. We identify EH galaxies above and below the Main Sequence of star formation (MS) with stars and triangles, respectively, following the definition of the MS from \cite{Schreiber2017}.}
         \label{fig:M_R}
   \end{figure}

\begin{figure}[!h]
   \centering
   \includegraphics[width=7cm]{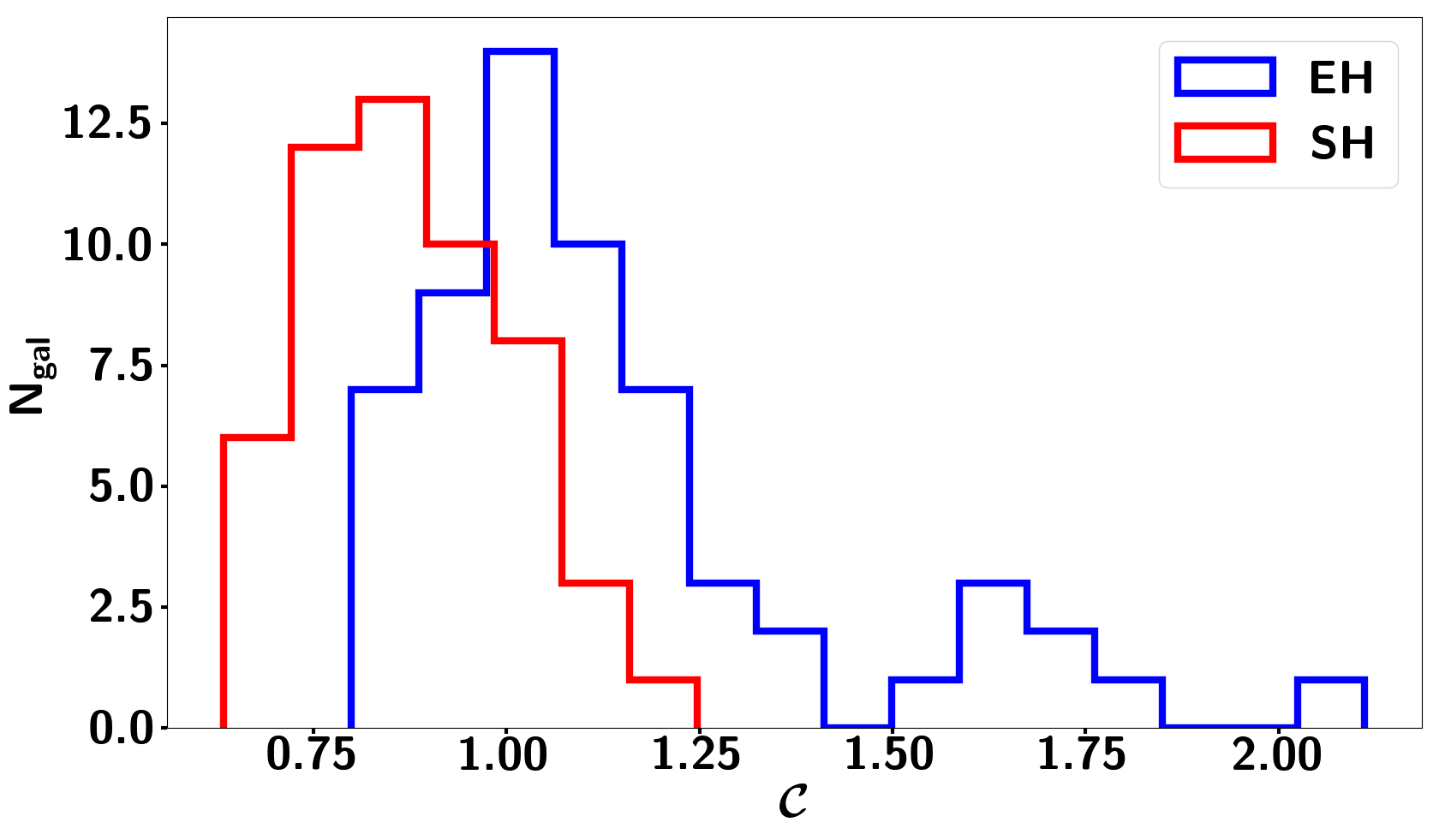}
      \caption{Compactness distributions for the EH and SH massive galaxies at $z=2$.}
         \label{fig:c_hist}
   \end{figure}

Stellar masses and radii are shown at $z=2$ in Fig.~\ref{fig:M_R}. 83\% of the galaxies in our sample are on the Main Sequence of star formation (MS, \cite{Elbaz2011}), so that we compare their size to the model from \citet{dutton-vdb11}, known to provide a good fit to MS galaxies at $z=2$\footnote{In the mass range studied here, the Dutton et al. model lies between the mass-size relations derived at $z\simeq 1.75$ and $z\simeq 2.25$ for SFGs in CANDELS by \citet{vanderwel}}. SH galaxies are larger than both EH galaxies and observed MS galaxies. EH galaxies generally lie around the observed relation, and a small fraction have significantly smaller sizes. We define the compactness $\mathcal{C}$ as the ratio between the radius expected from the \cite{dutton-vdb11} model and the actual radius. The compactness distribution for EH (Fig.~\ref{fig:c_hist}) peaks at around $\mathcal{C} \simeq 1$ but exhibits a distinct tail for $\mathcal{C} > 1.3$. We thus define two massive galaxy populations in EH: 10 ultra-compact (UC) galaxies with $\mathcal{C} > 1.3$ and 50 non ultra-compact (NUC) ones.

Hence, massive galaxies in EH are globally more compact than in SH, and EH contains a population of UC outliers. The larger sizes in SH do not just correspond to extended stellar haloes: the difference remains when we vary the surface density threshold in mock images, and Sersic indices are on average similar in EH and SH. The size difference is not expected to arise from internal processes such as instabilities and/or feedback, as galactic scales and feedback are treated with the very same resolution in EH and SH. Two key differences could contribute: EH models gas flows in the CGM at a much higher resolution, and low-mass galaxies are under-resolved in SH.

\subsection{Diffuse accretion and angular momentum supply}
\label{sec:momentum}

A substantial part of the angular momentum of galaxies is supplied by cold gas inflows \citep{ocvick08,Pichon2011,danovich, Tillson2015} which are better resolved in EH. Higher resolution could also better probe metal mixing in the IGM and subsequent cooling \citep{pichon}. To probe these potential effects, we focus on inflowing gas in the vicinity of massive galaxies using the following criteria, which typically select inflowing gas according to other simulations \citep[e.g.,][]{goerdt}: 
\begin{itemize}
    \item a galactocentric radius between $3\,R_e$ and 50\,kpc, 
    \item a density below 0.1\,cm$^{-3}$ to exclude satellites, 
    \item a velocity vector pointing inwards w.r.t. the galaxy center,
    \item a temperature below $10^{5.5}$\,K.
\end{itemize}

For each resolution element following these criteria, we compute the gas mass $m$ and angular momentum $l$ w.r.t the galaxy center (in norm, $l=\|  \Vec{l} \|$), sum-up the total angular momentum $L=\mathrm{\Sigma}\,l $ and mass $M=\mathrm{\Sigma}\,m $ for inflowing gas, and compute the specific momentum of inflowing gas $\mathcal{L}_{in}=L/M$ around each galaxy. Differences in $\mathcal{L}_{in}$ for various galaxy samples are listed in Table~\ref{tab:L}, showing that $\mathcal{L}_{in}$ around massive galaxies is substantially lower in EH than in SH, but is almost similar around EH-UC and EH-NUC galaxies.

  \begin{table}
  \begin{tabular}{lc}
   \hline \hline
Galaxy samples  &  Mean difference in $\mathcal{L}_{in}$\\
  \hline
EH vs. SH  &    $13$\% lower \\
EH-NUC vs. SH  &    $10$\% lower \\
EH vs. SH at $M_* < 10^{11}$\,M$_\sun$  &    $12$\% lower \\
EH-UC vs. EH-NUC at $M_* < 10^{11}$\,M$_\sun$  &    $3$\% lower \\
     \hline
 \end{tabular}
 \caption{Mean difference in the specific angular momentum of inflowing gas $\mathcal{L}_{in}$ between several samples of massive galaxies.}
  \label{tab:L}
 \end{table}

   \begin{figure*}[!h]
   \centering
   \includegraphics[height=3.8cm]{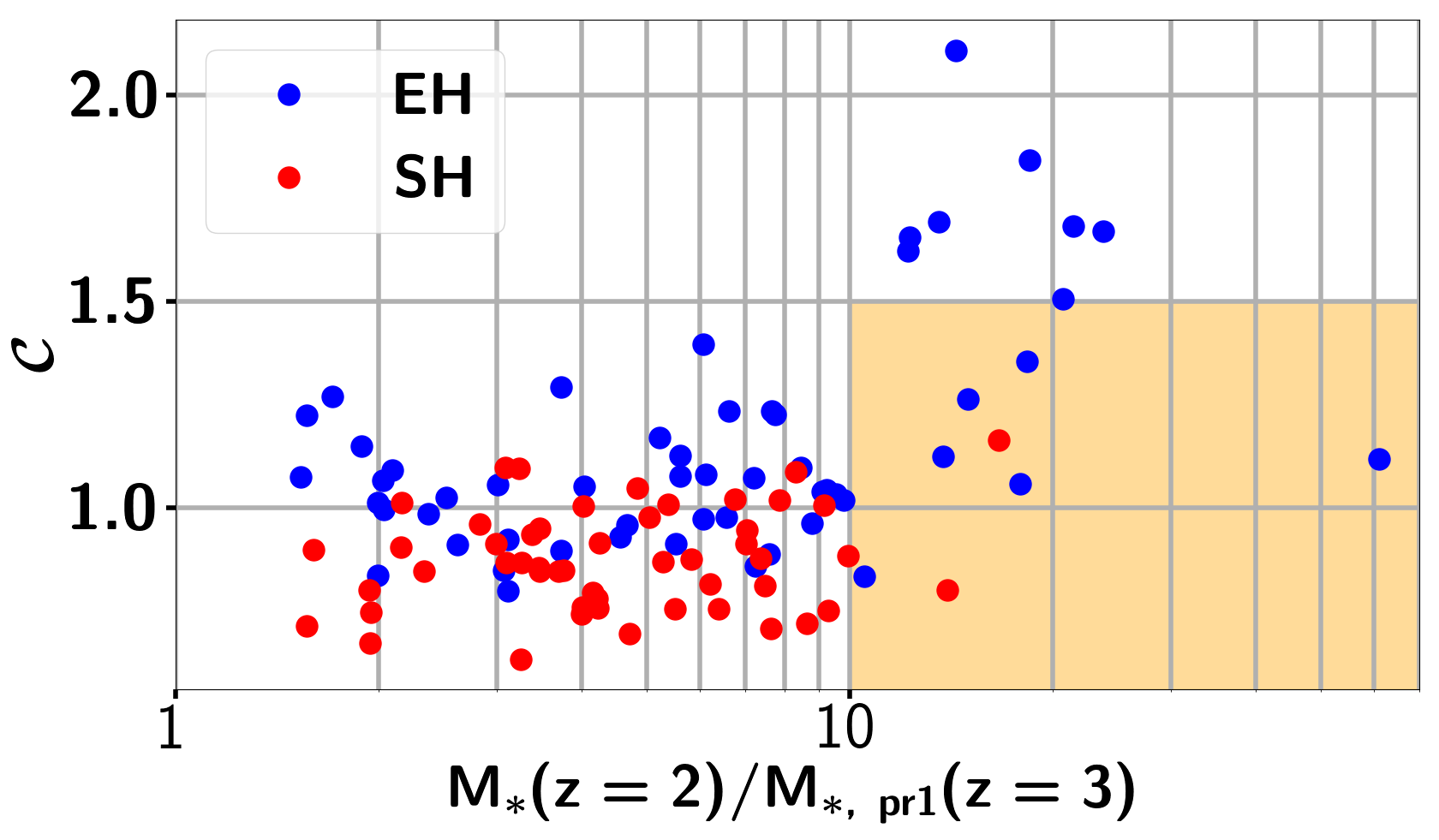} \hspace{-1.8mm}
   \includegraphics[height=3.8cm]{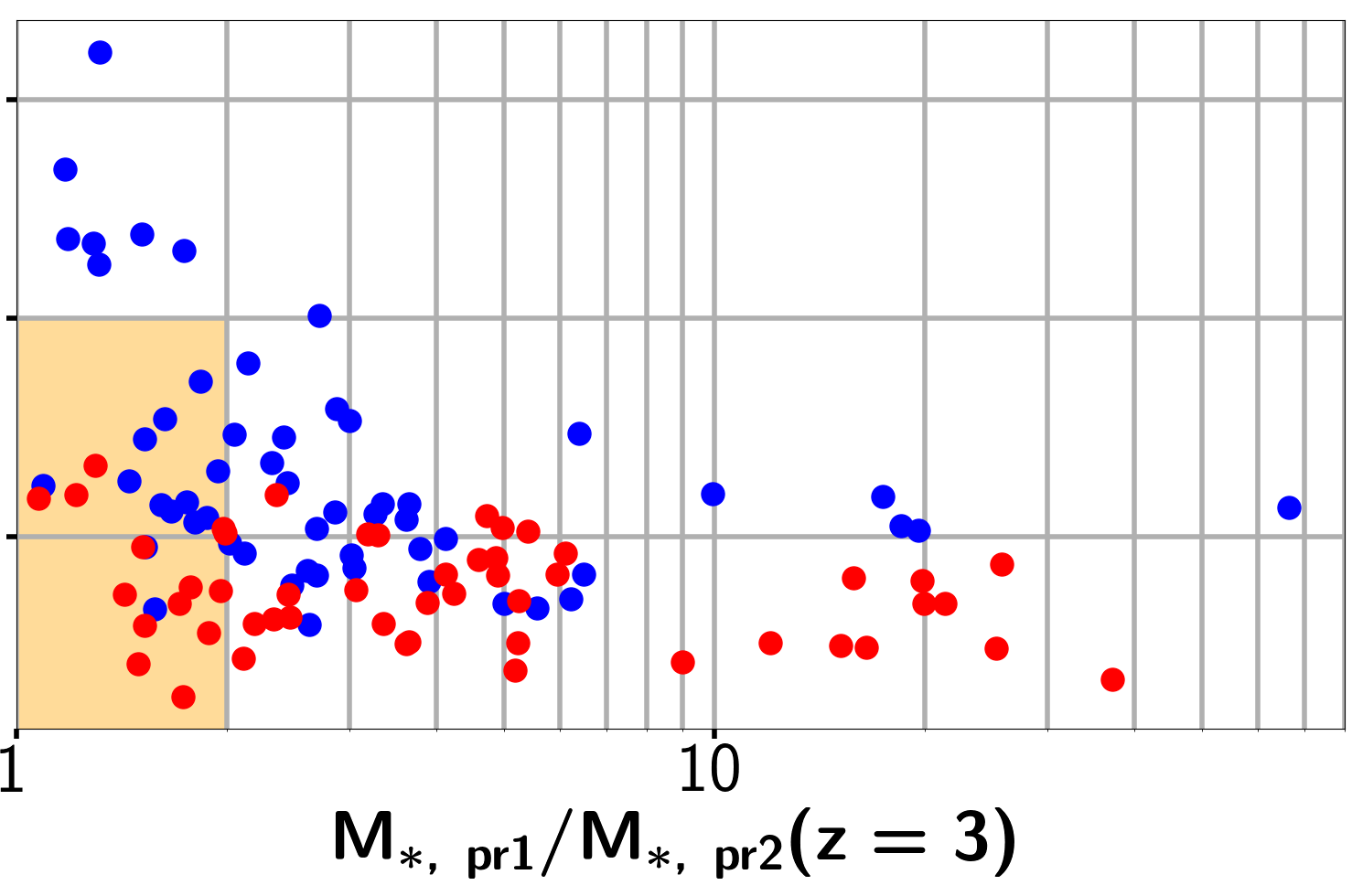}
   \includegraphics[height=3.8cm]{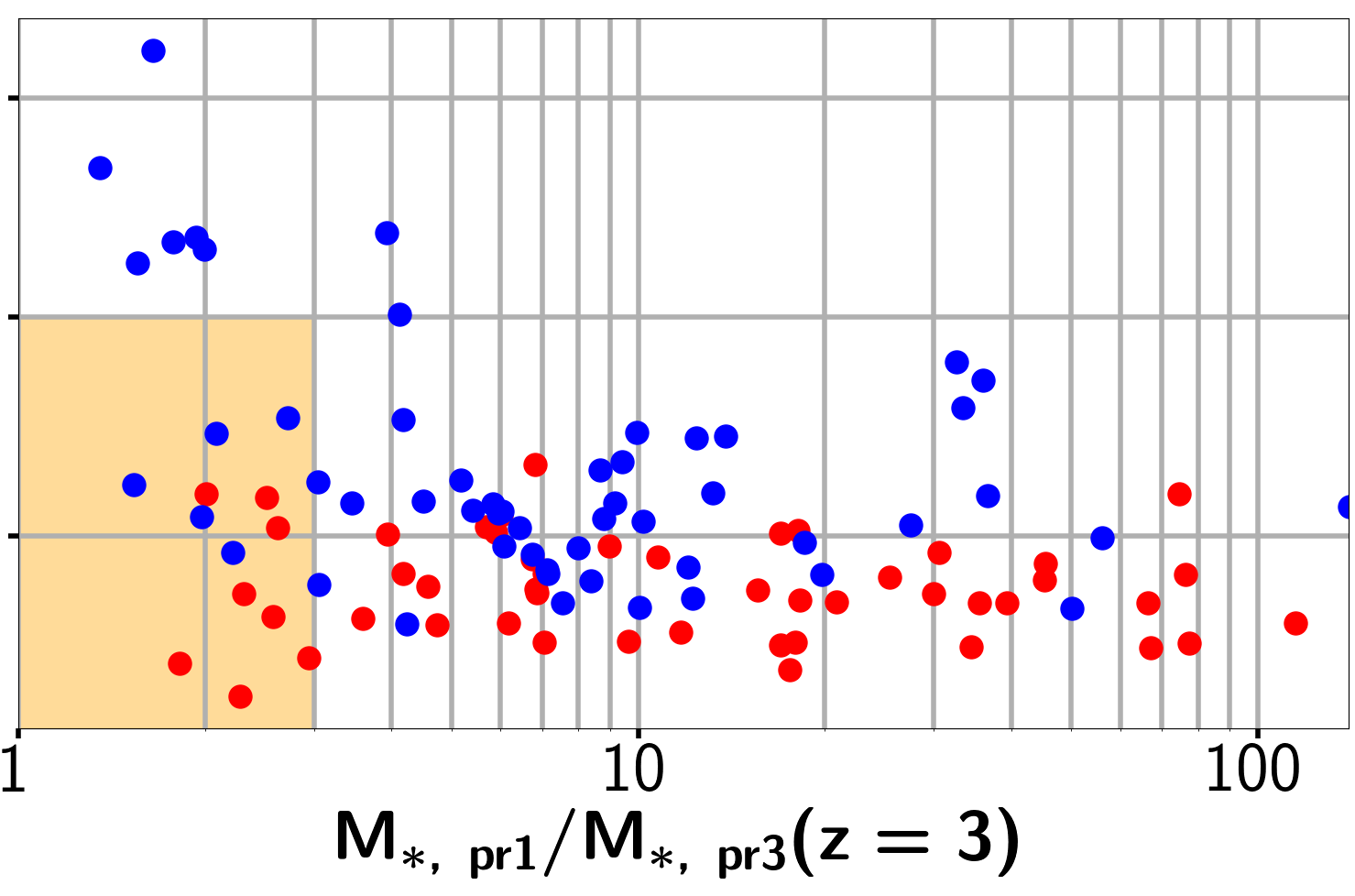}
      \caption{Compactness $\mathcal{C}$ as a function of the ratio between the stellar mass at $z=2$, $M_{\rm *} (z = 2)$, and the mass of the main progenitor at $z=3$ $M_{*\rm , \  pr1} (z = 3)$  (left panel) and as a function of the ratios between the mass of the three most massive progenitors (first  and second, $M_{*\rm , \  pr2}$ on  the middle panel and first and third, $M_{*\rm , \  pr3}$, on the right panel) for EH (blue) and SH (red) galaxies. The shaded areas define regions of galaxies that would grow through major mergers of low-mass galaxies but would end-up NUC: no SH galaxy and only one EH galaxy are in the three areas at the same time.}
         \label{fig:c_minprog}
   \end{figure*}

We can estimate the potential impact on galaxy sizes under two extreme assumptions. On the one hand, if the circular velocity remains unchanged, dominated by a non-contracting DM halo, then galactic radii should follow $R \propto \mathcal{L}_{in}$. On the other hand, if the dark matter halo contracts in the same proportions as the baryons, the rotation velocity $V$ and radius $R$ follow $V^2 \propto 1/R$ at fixed mass, so that $R \propto \mathcal{L}_{in}^2$. 

Hence the 10\% difference in $\mathcal{L}_{in}$ between EH-NUC and SH could result in a 10--20\% size difference: this can account for the smaller sizes of massive galaxies in EH compared to SH. On the other hand, the population of UC galaxies does not result from diffuse gas accretion as it could only impact sizes by a few percent compared to NUC galaxies.

\medskip

Angular momentum is built up by tidal torques that only depend on very large-scale structures expected to be well resolved even in SH \citep{FE80}. Yet, angular momentum can be lost when cold inflowing streams interact with hot gas haloes and outflows in the CGM. Idealized simulations of cold streams interacting with hot haloes \citep{Mandelker20} indicate that instabilities can decrease the velocity of cold streams by up to a few tens of percent in favorable cases, which can explain the loss of angular momentum at the EH resolution compared to SH\footnote{Mandelker et al. suggest that 10--20 resolution elements per stream diameter are required to model such instabilities. For our typical filament diameter of 20--30\,kpc at $z$=2--3, EH reaches such resolution in the CGM, but SH does not (Appendix~A, Fig.~A2).}.

\subsection{Major mergers of low-mass progenitors}
\label{sec:mergers}

Another driver of compaction could be the numerous low-mass galaxies in EH that are missing in SH. We identify the progenitors of $z=2$ UC and NUC galaxies by tracking their stellar particles, and analyze their progenitors at $z=3$ and $z=4$ with the same technique as our $z=2$ sample.

Figure.~\ref{fig:c_minprog} shows the compactness as a function of the mass ratios between each $z=2$ galaxy and its main $z=3$ progenitor and between the main $z=3$ progenitor and the second and third most massive progenitors. UC galaxies have (1) a main $z=3$ progenitor that never exceeds 10\% of the $z = 2$ mass, (2) a second and (3) third most massive progenitors almost as massive as the main progenitor, with mass ratios lower than 3:1 (generally lower than 2:1) for the second most massive, and generally below 4:1 for the third most massive. This pinpoints a correlation between these parameters, showing that the formation of EH-UC galaxies involves repeated\footnote{similar criteria hold for the fourth and fifth most massive progenitors and are also valid when the same analysis is performed at $z=4$.} major mergers between low-mass progenitors. These mergers occur rapidly between $z = 3$ and $z = 2$ with 80\% of UC galaxies that assemble 90\% of their stellar mass in this redshift range. Conversely, 70\% of galaxies that have assembled 90\% of their stellar mass between $z=3$ and $z=2$ end up as UC galaxies.

In contrast, EH-NUC and SH galaxies most often have one dominant progenitor undergoing only minor mergers, and very rarely meet the three criteria depicted above for UC formation at the same time. There is actually no SH galaxy and only one EH-NUC galaxy that lies in the three shaded areas in Fig.~\ref{fig:c_minprog} at the same time. This strengthens our argument that these specific types of accretion histories essentially always produce UC galaxies. The only exception among EH-NUC galaxies has an extended spiral disk morphology, and has the second highest total angular momentum $L$ in inflowing gas over the whole EH sample so that accretion of diffuse gas  compensates for the compacting effects of the merger history in this extreme object. It is expected from idealized simulations of repeated mergers with various mass ratios that mergers histories involving mostly major mergers with relatively similar masses produce more concentrated end-products for the same total merged mass (at least in terms of Sersic indices, \citealt{B07}, Fig.~4). 45\% and 47\% of the stars found in EH-NUC and SH galaxies at $z = 2$ are already formed at $z=3$, respectively, compared to only 36\% for EH-UC galaxies: UC galaxies arise from low-mass progenitors and hence form their stars later on.
 
We also note that the distributions of progenitor masses are fairly identical for EH-NUC and SH galaxies (Fig.~\ref{fig:c_minprog}) indicating that the smaller sizes of EH-NUC galaxies doe not result from different merger histories but rather from the modeling of diffuse gas infall (Sect.~3.3).

%-------------------------------------------------------------------
\section{Discussion}
\label{sec:discussion}

In order to match the resolution of SH and HAGN in galaxies, the EH simulation is limited to kpc-scale resolution, so the real compactness of UC galaxies could be under-estimated as they are as compact as the resolution limit allows. Zoom-in simulations will be required to make robust assessment of their size distribution. Nevertheless, the population of UC galaxies in EH is tightly associated with specific formation histories dominated by major mergers of low-mass progenitors, compared to larger galaxies in the simulation. 

To further probe the effect of feedback in compact galaxy formation, we used the Horizon-AGN suite of simulations from~\citet{Chabanier2020}. These simulations are run with extreme feedback parameters leading to barely realistic variations of the black hole-to-stellar mass ratio, yet the average galaxy size at fixed stellar mass changes by less than 10\%, confirming that feedback is not a key driver of the formation of UC galaxies in EH.

\medskip

   \begin{figure}[!h]
   \centering
   \includegraphics[width=\columnwidth]{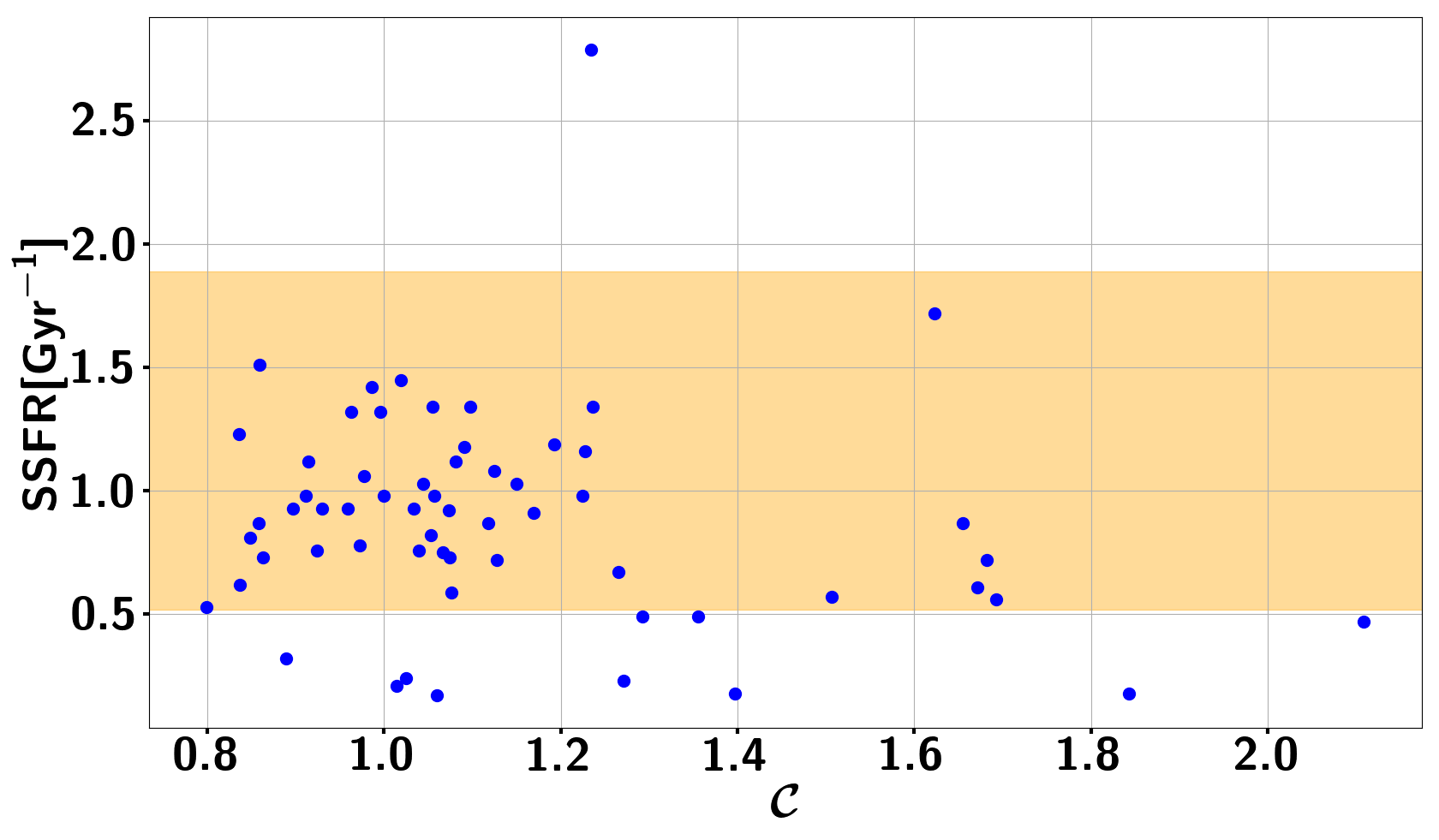}
      \caption{Specific Star Formation Rate (SSFR) as a function of compactness ($\mathcal{C}$ for EH galaxies at $z=2$. The shaded area defines the Main Sequence following \cite{Schreiber2017}.}
         \label{fig:ssfr}
   \end{figure}

We have analyzed so far the compactness of galaxies independently from their star formation activity. As expected for galaxies in the $10^{10}$-$10^{11}$\,M$_\sun$ stellar mass range at $z=2$, both NUC and UC galaxies are mainly star-forming galaxies on the MS. There is nevertheless a clear trend for compact galaxies to have relatively low specific star formation rates (sSFR, Fig.~\ref{fig:ssfr}). The majority of UC galaxies lie on the low-sSFR end of the MS, as observed for blue nuggets \citep{barro17}. The relatively low sSFRs of UCs, as well as a tentative excess of galaxies below the MS among UCs compared to NUCs, are consistent with the idea that these objects are undergoing quenching through gas exhaustion and/or feedback \citep{tachella}. 

The number of UC galaxies in EH (10 objects in (50\,Mpc/h)$^3$) is consistent with the number density of compact SFGs (see Introduction). The EH volume is too small to firmly probe the formation of massive compact ETGs at $z$=2, as statistically about one such object is expected in this volume, but the excess of low-mass progenitors in EH is already present at $z$=4 (Fig.~\ref{fig:Mgal}) and could explain the early formation of such compact ETGs. There is indeed one galaxy in EH with $M_*$=1.2$\times$ $10^{11}$\,M$_\sun$ and compactness $\mathcal{C}$=1.29 (almost UC in our definition), with a low SSFR=0.23\,Gyr$^{-1}$ (a factor 7 below the MS), a low gas fraction of 11\% (within 3$R_e$), and a Sersic index of 3.6 at $z$=2. This galaxy continues to quench into a compact ETG by redshift $z \simeq 1.8$, with SSFR=0.13\,Gyr$^{-1}$, $M_*$=1.7$\times$ $10^{11}$\,M$_\sun$, and $R_e$=4.0\,kpc at $z \simeq 1.8$, thus lying close to the mass-size relation of ETGs at $z=1.75$ from \citet{vanderwel}. This candidate compact ETG does also form through major mergers of low-mass progenitors: its two main progenitors at $z$=4 contain 11 and 8\% of its stellar mass, respectively.

We also examined the environment of UC and NUC galaxies in EH by studying the large-scale structure with the persistent skeleton approach \citep{2011MNRAS.414..350S}. UC galaxies are found in relatively dense environments, but not in the very densest filaments and nodes (see Appendix~\ref{app:skeleton}). This strengthens our previous findings on the merger history of UC galaxies, as objects in the densest regions of the main filaments are expected to form their main progenitor early-on and subsequently grow by minor mergers and diffuse accretion.

\section{Conclusion}

 In this Letter, we introduced the EH cosmological hydrodynamical simulation, based on the physical model of HAGN, with a substantial increase in the spatial resolution in the IGM and CGM while galactic scales are treated at the same resolution. The SH simulation of the same volume uses a lower resolution in the CGM and IGM, more typical in cosmological simulations.

The comparison of the mass-size relation of massive galaxies in EH and SH  highlights the importance of modeling diffuse gas flows at high-enough resolution in the IGM and CGM, as this tends to reduce the angular momentum supply onto massive galaxies. In addition, the EH simulation produces a population of ultracompact (UC) galaxies. These form rapidly by repeated major mergers of low-mass progenitors, which can be missed in simulations using a modest resolution in low-density regions. A pleasant outcome of our analysis is that issues in galaxy formation simulations could indeed be solved by accurately resolving structure formation without calling upon feedback or novel subgrid models.%-------------------------------------------------------------------
\begin{acknowledgements}
 The Extreme-Horizon simulation was performed as a ``Grand Challenge'' project granted by GENCI on the AMD Rome extension of the Joliot Curie supercomputer at TGCC. We are indebted to Marc Joos, Adrien Cotte, Christine M\'enach\'e and the whole HPC Application Team at TGCC for their efficient support. Collaborations and discussions with Bruno Thooris, Eric Armengaud, Marta Volonteri, Avishai Dekel, are warmly acknowledged. We deeply appreciate comments from J\'er\'emy Blaizot on the Extreme Horizon project. This research used the {\sc ramses} code written mainly by Romain Teyssier, the custom {\sc Hercule}  parallel I/O library, the {\sc kinemetry} package written by Davor Krajnovi\'c, and the {\sc disperse} code from Thierry Sousbie. This work was supported by the ANR 3DGasFlows (ANR-17-CE31-0017) and made use of the Horizon Cluster hosted by the Institut d'Astrophysique de Paris, run by St\' ephane Rouberol. SCo's research is partially supported by Fondation MERAC. TK was supported by the National Research Foundation of Korea (NRF-2017R1A5A1070354 and NRF-2020R1C1C100707911). MT is supported by Deutsche Forschungsgemeinschaft (DFG, German Research Foundation) under Germany's Excellence Strategy EXC-2181/1 - 390900948 (the Heidelberg STRUCTURES Cluster of Excellence).
 
\end{acknowledgements}

% WARNING
%-------------------------------------------------------------------
% Please note that we have included the references to the file aa.dem in
% order to compile it, but we ask you to:
%
% - use BibTeX with the regular commands:
   \bibliographystyle{aa} % style aa.bst
   \bibliography{biblio} % your references Yourfile.bib
%
% - join the .bib files when you upload your source files
%-------------------------------------------------------------------

%
\begin{appendix} 
\section{Overview of the EH simulation} \label{app:EH}

\begin{figure*}[!h]
   \centering

  \includegraphics[width=\textwidth]{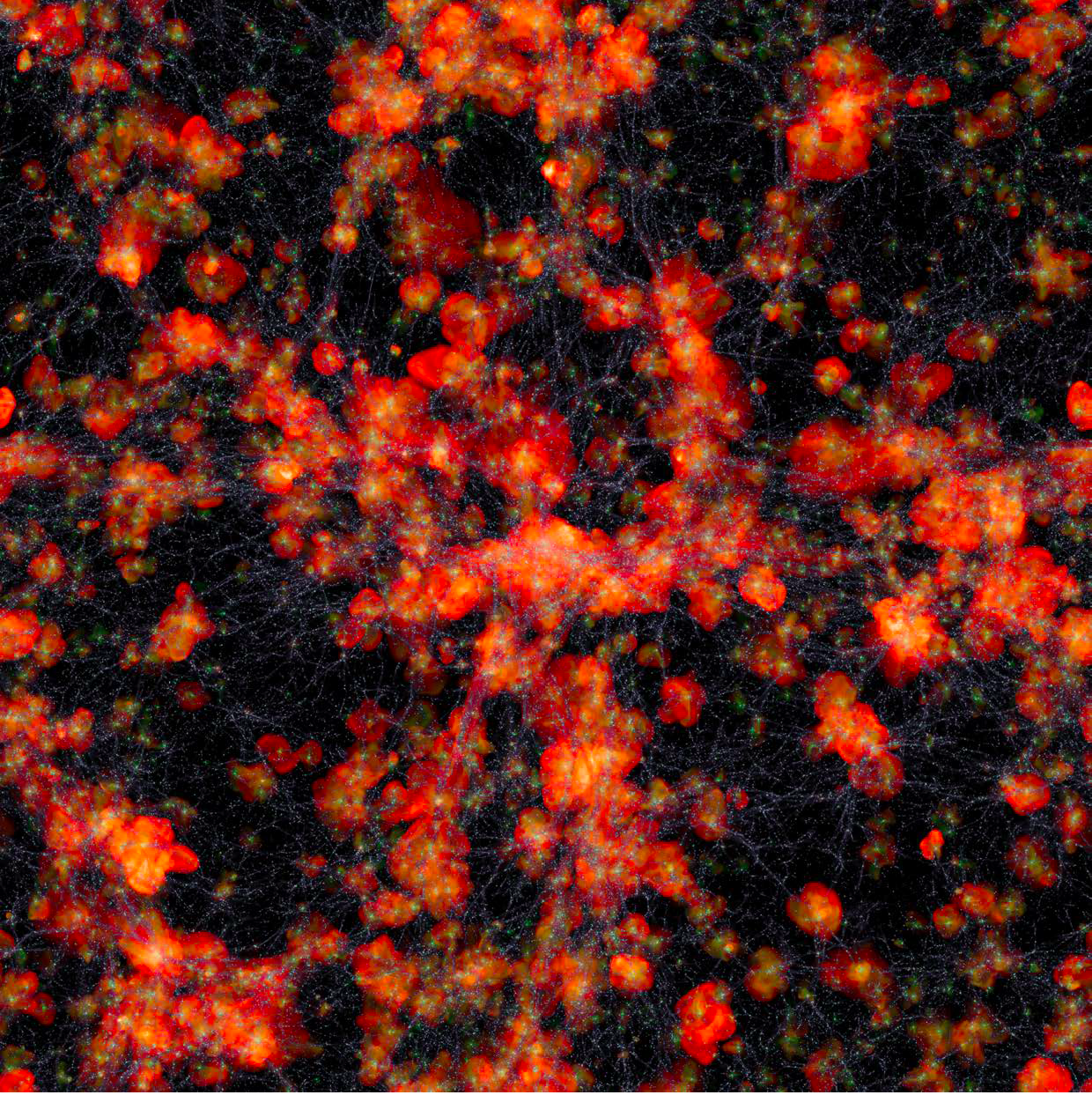}
      \caption{Projected map of the EH simulation at $z\simeq 2$. Gas density (grey), entropy (red) and metallicity (green) are shown. }\label{fig:a1}
   \end{figure*}

\begin{figure*}[!h]
   \centering
   \includegraphics[width=\textwidth]{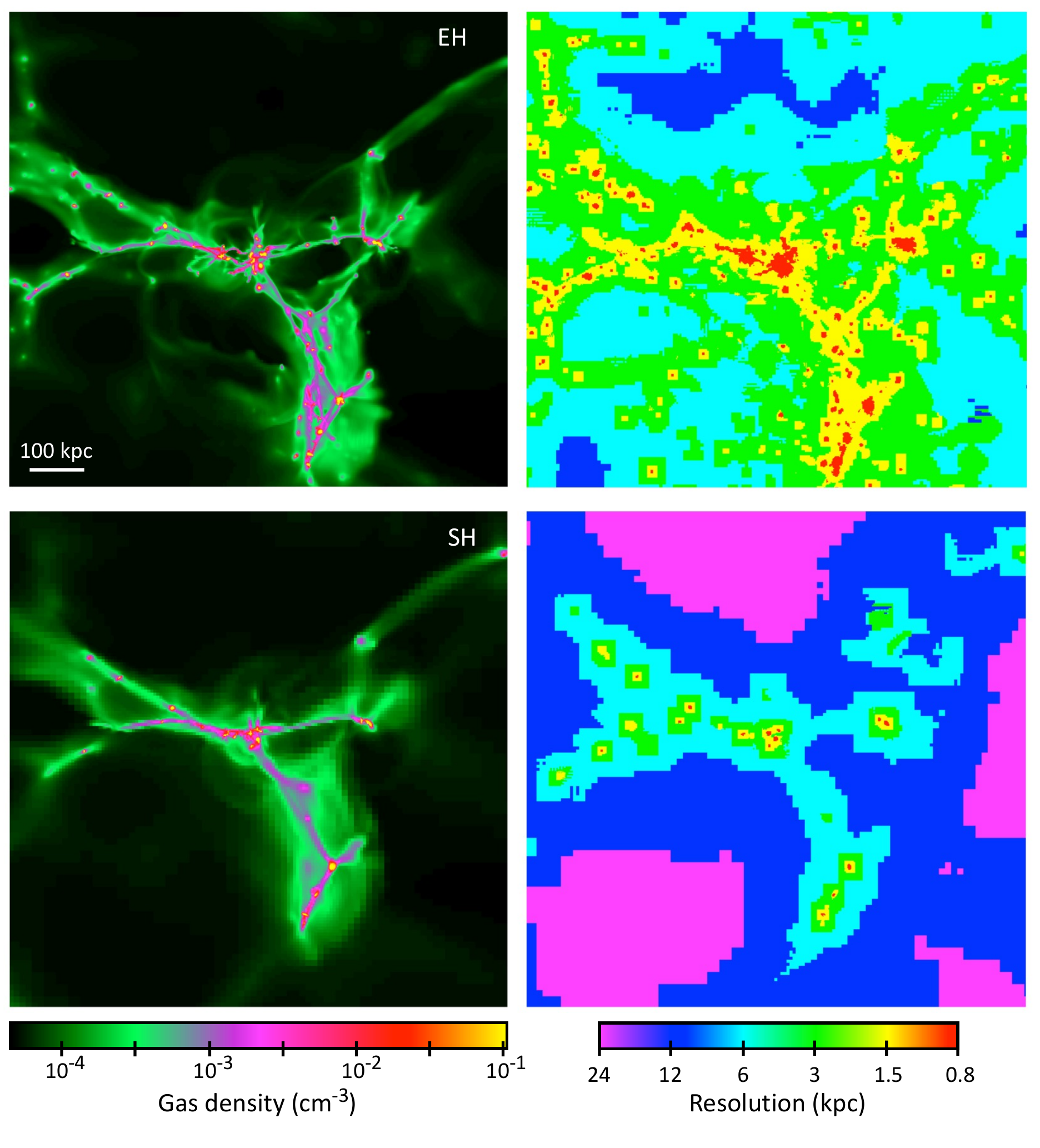}
      \caption{Projected density (left) and physical resolution (right) in EH (top) and SH (bottom) zoomed on a massive halo at $z = 3$. The depth of the projections are 200\,kpc$/h$ and the boxes extend 1\,Mpc$/h$ on each side. The gas density is computed as the mass-weighted average of local densities along the line-of-sight corresponding to each pixel. The resolution shown is the resolution of the cell in which the gas density is the highest along each line-of-sight.}\label{fig:a2}
   \end{figure*}

Fig.~\ref{fig:a1} shows the large-scale structure of the EH simulation at redshift $z=2$. Fig.~\ref{fig:a2} displays the gas density in the CGM and IGM around a massive halo along with the spatial resolution achieved in the EH and SH simulation in the same region.

\section{Massive galaxies in EH and SH} \label{app:maps}

\begin{figure*}[!h]
   \centering
   \includegraphics[width=15cm]{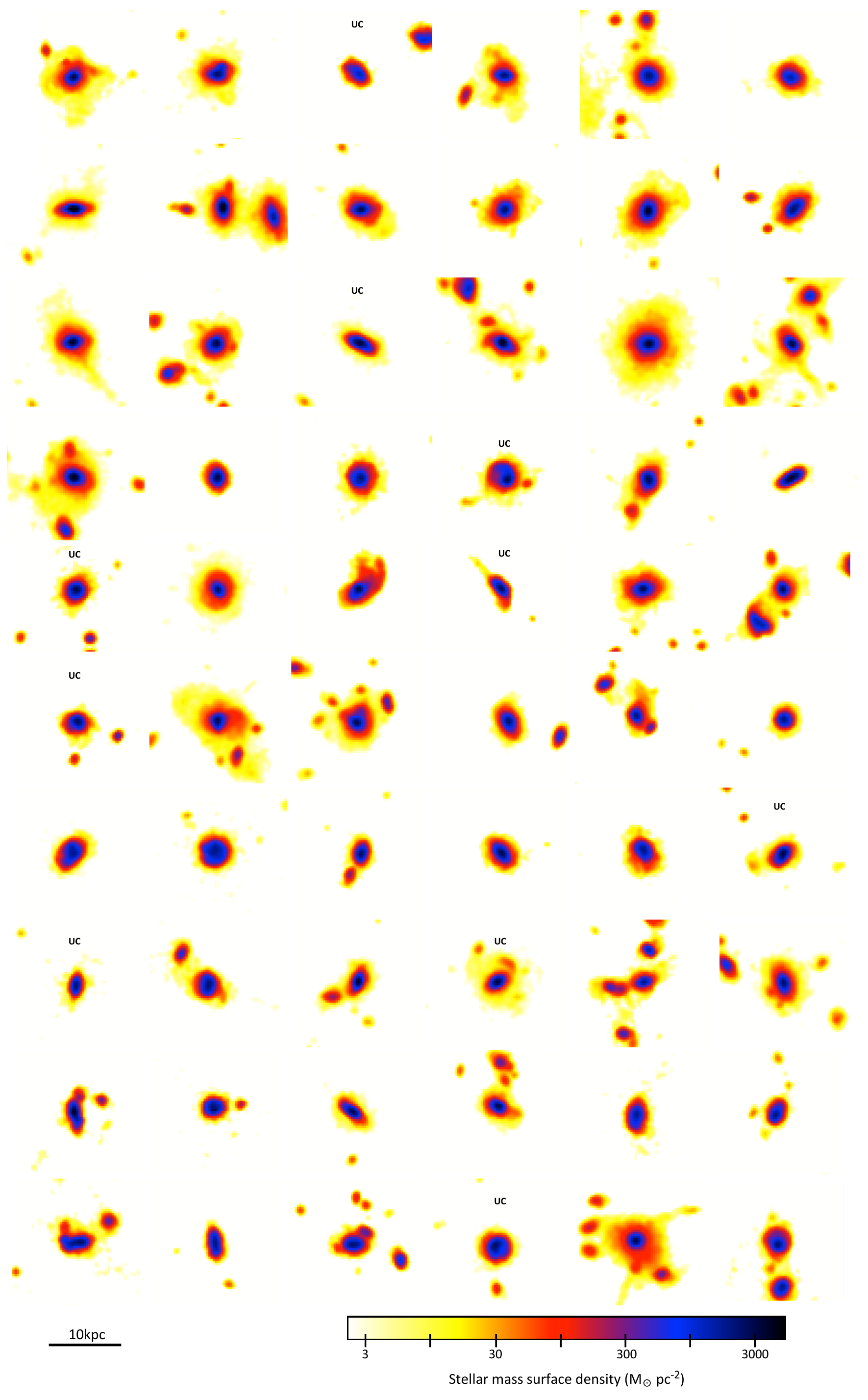}
      \caption{Stellar mass distribution of massive galaxies in the EH simulation. UC galaxies are flagged.}
      \label{fig:b1}
   \end{figure*}

\begin{figure*}[!h]
   \centering
   \includegraphics[width=15cm]{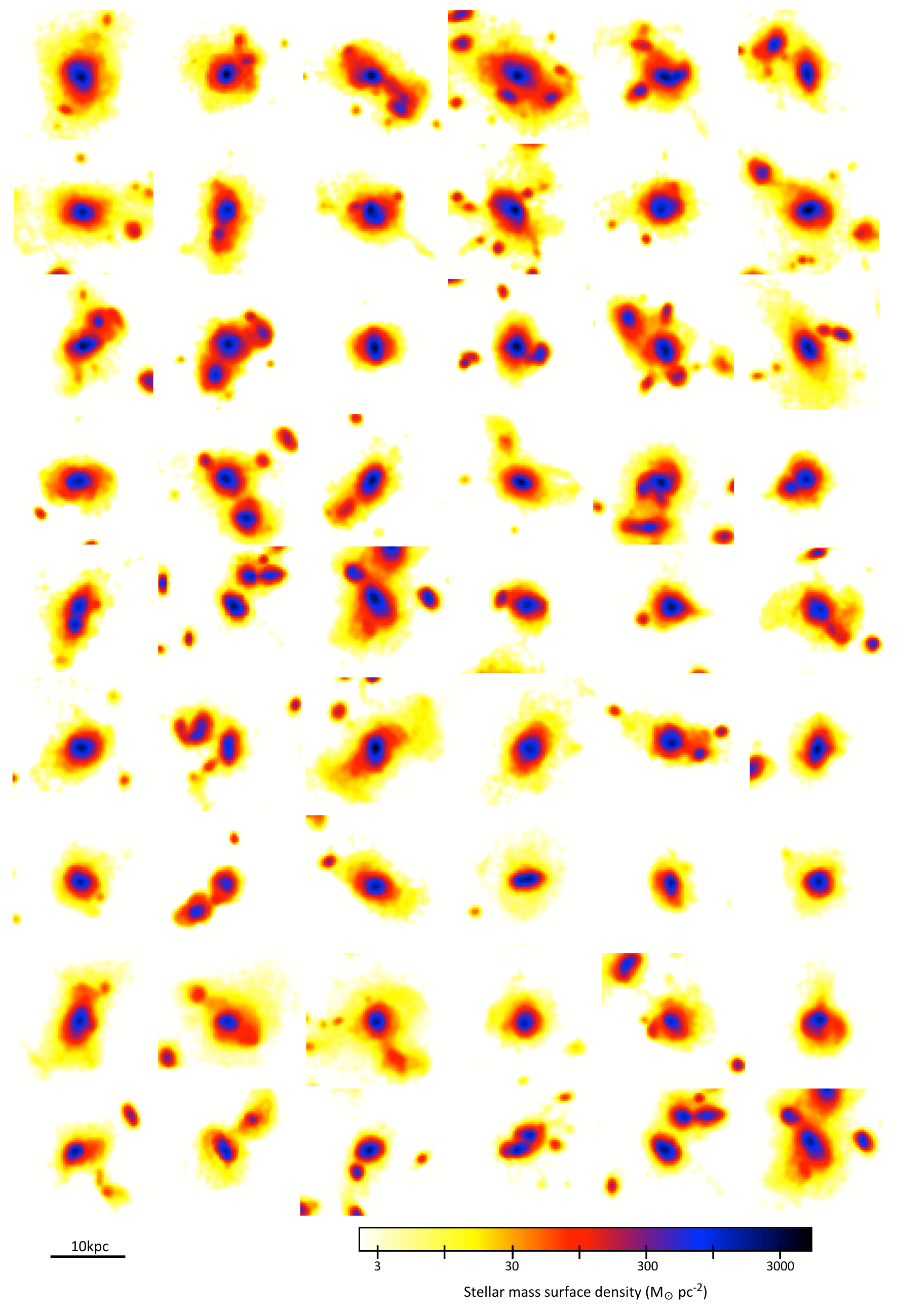}
      \caption{Stellar mass distribution of massive galaxies in the SH simulation. Galaxies are not meant to be individually matched to SH galaxy as independent samples were built in EH and SH.}
      \label{fig:b2}
   \end{figure*}

Galaxy stellar mass maps from EH and SH are shown in Fig.~\ref{fig:b1} and Fig.~\ref{fig:b2} respectively. The slightly smaller sample size in SH compared to EH results from major mergers that do not occur at the very same time in both simulations and from a few galaxies that are just below the mass cut-off in SH.

\section{Environmental dependence}
\label{app:skeleton}

\begin{figure*}[!h]
   \centering
   \includegraphics[width=15cm]{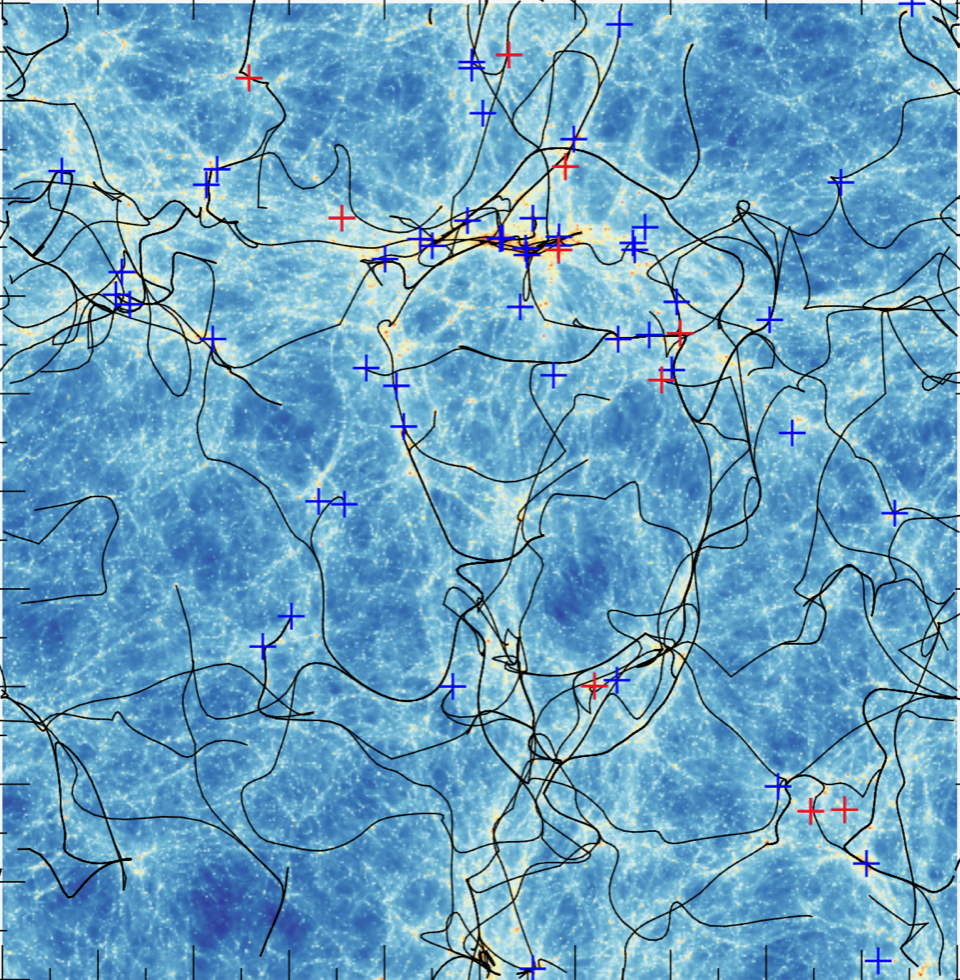}
      \caption{Projected EH skeleton of the 50 Mpc h$^{-1}$ box with a $7\sigma$ persistence level at $z = 2$. Crosses indicate the projected position of massive galaxies, with red crosses for UC and blue crosses for NUC.}
      \label{fig:skeleton}
   \end{figure*}

To compare the environment of UC and NUC galaxies, we study the large-scale structure of the EH simulation with the persistent skeleton approach \citep{2011MNRAS.414..350S} using the {\tt DISPERSE} code \citep{Sousbie2013}. The full skeleton is shown in Fig.~\ref{fig:skeleton}.  Topological persistence can be used to characterise the significance of the structures depending on the local level of noise. Persistence levels from $3$ to $8\,\sigma$ are used to investigate different scales and prominences of the corresponding cosmic web. 

\begin{figure}[!h]
   \centering
   \includegraphics[width=\columnwidth]{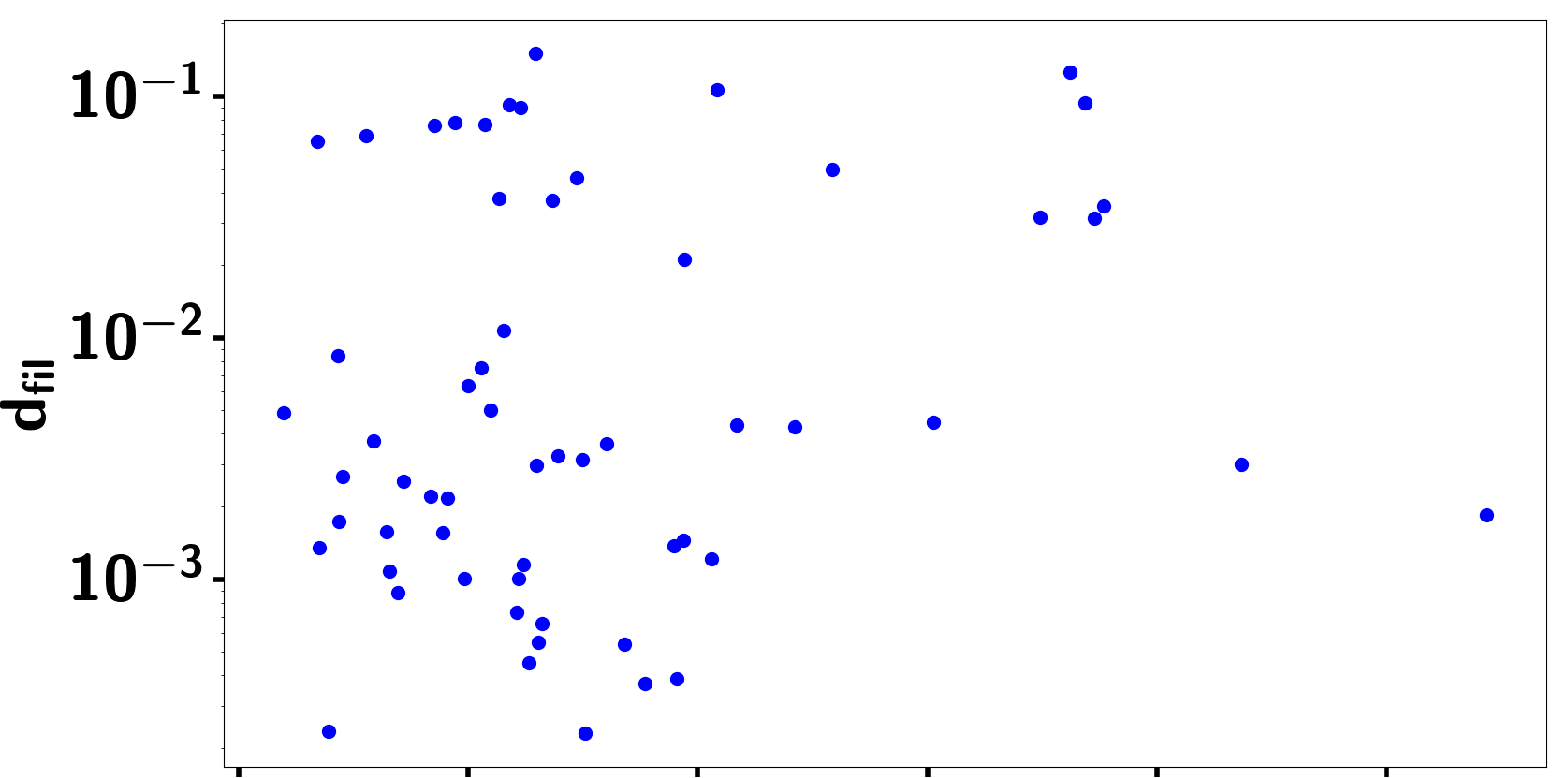}
   \includegraphics[width=\columnwidth]{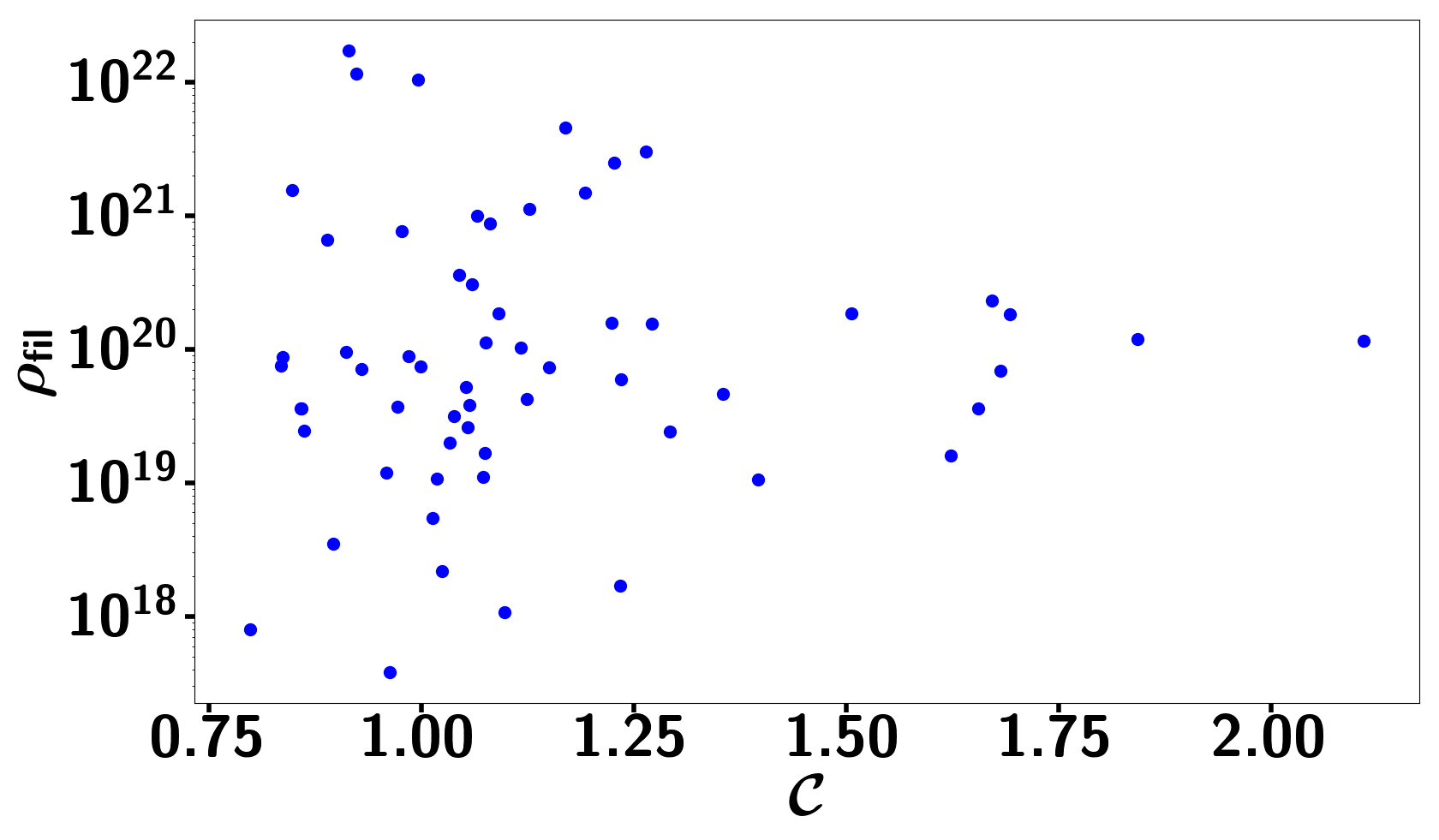}
      \caption{\textbf{Top panel:} Distance to the closest filament $d_{\rm fil}$ (in box size units) of the 8$\sigma$ sparse skeleton for all EH massive galaxies as a function of their compactness. \textbf{Bottom panel} Density in the closest filament $\rho_{\rm fil}$ (obtained by DTFE from a mass-weighted Delaunay tessellation of the galaxy catalogue) of the 3$\sigma$ dense skeleton for all EH massive galaxies as a function of their compactness. For UC galaxies, exclusion zones are clearly visible at small distance to the filaments and in the very low and very high density regions, compared to the NUC.}
         \label{fig:skl}
   \end{figure}

At high persistence, the skeleton is sparse, dominated by a few dense and extended filaments. UC and NUC galaxies both lie close to such filaments, as expected for massive galaxies in general, but the galaxies that lie closest to these dense filaments and their nodes are never UC (Fig.~\ref{fig:skl}, top panel). Instead, UC galaxies tend to lie in intermediate-density filaments, as shown by the analysis of the closest filaments in a lower-persistence skeleton analysis (Fig.~\ref{fig:skl}, bottom panel). 
This is consistent with the previous results on the merger history of UC galaxies, as objects in the densest regions of the main filaments are expected to form their main progenitor early-on and subsequently grow by minor mergers and/or diffuse accretion. UC galaxies nevertheless still do form in dense regions and none is found in low-density filaments where smooth accretion would dominate over mergers (Fig.~\ref{fig:skl} right panel and Fig.~\ref{fig:skeleton} for a visualisation).

Hence, UC galaxies are expected to be found in relatively dense environments, but not in the very densest filaments and nodes. Galaxies in the densest regions of the cosmic web are expected to be rarely ultra-compact at $z \sim 2$, yet could undergo ultra-compact phases at higher redshift if their early formation involves major mergers of numerous low-mass progenitors.

\end{appendix}

\end{document}